# PROTOTYPE FOR EXTENDED XDB USING WIKI

Wook-Sung Yoo[1]

[1]Department of Software Engineering, Fairfield University, Fairfield, CT, USA
wyoo@fairfield.edu

## ABSTRACT

*This paper describes a prototype of extended XDB. XDB is an open-source and extensible database architecture developed by National Aeronautics and Space Administration (NASA) to provide integration of heterogeneous and distributed information resources for scientific and engineering applications. XDB enables an unlimited number of desktops and distributed information sources to be linked seamlessly and efficiently into an information grid using Data Access and Retrieval Composition (DARC) protocol which provides a contextual search and retrieval capability useful for lightweight web applications. This paper shows the usage of XDB on common data management in the enterprise without burdening users and application developers with unnecessary complexity and formal schemas. Supported by NASA Ames Research Center through NASA Exploration System Mission Directorate (ESMD) Higher Education grant, a project team at Fairfield University extended this concept and developed an extended XDB protocol and a prototype providing text-searches for Wiki. The technical specification of the protocol was posted to Source Forge (sourceforge.net) and a prototype providing text-searches for Wiki was developed. The prototype was created for 16 tags of the MediaWiki dialect. As part of future works, the prototype will be further extended to the complete Wiki markups and other dialects of Wiki.*

## 1. INTRODUCTION

Searching, extracting, and integrating information from documents is a key requirement in many enterprise-wide information systems applications [1]. However, it is estimated that 80% of all corporate information is unstructured or "semi-structured" (i.e. some structure in the documents but not exactly a formal structure) [2]. The engineering teams in National Aeronautics and Space Administration (NASA) have struggled to access information in such semi-structured documents with hundreds of different formats in their own explicit and implicit structures. The decision making applications to access this information are required to follow numerous procedures, guidelines, and complex work practices. To resolve this problem, XDB, an open-source and extensible database architecture, was developed by NASA Ames Research Center to provide seamless integration of these heterogeneous and distributed information resources for scientific and engineering applications [3, 4]. To prove the basic concept of XDB protocol, which was currently implemented for documents in XML and HTML formats only, a software engineering program team at Fairfield University worked with NASA Ames Research Center to extend the XDB for different document formats. As a part of the NASA Exploration System Mission Directorate's Higher Education Project, a draft of the XDB protocol and source code were provided to the Fairfield team to create a prototype extending XDB to Wiki document, a new document format.

The paper is organized as follows. Section 2 discuss about related research. Section 3 briefly reviews current XDB system and describes the prototype of extended XDB developed for Wiki document format. Section 4 concludes the paper and provides future work.

## 2. RELATED RESEARCH

XDB provides a novel "schema-less" database approach using a document-centered object-relational XML database mapping [5, 6]. This enables structured, unstructured, and semi-structured information to be integrated without requiring document schemas or translation tables. XDB utilizes existing international protocol standards of the World Wide Web Consortium Architecture Domain and the Internet Engineering Task Force, primarily HTTP, XML, and WebDAV (Web Distributed Authoring and Versioning), an extension of the Hypertext Transfer Protocol (HTTP) that facilitates collaboration between users in editing and managing documents and files stored on World Wide Web servers [7, 8, 9, 10]. Through a combination of these international protocols, universal database record identifiers, and physical address data types, XDB enables an unlimited number of desktops and distributed information sources to be linked seamlessly and efficiently into an information grid [11]. XDB uses Data Access and Retrieval Composition (DARC) protocol, a REST HTTP query protocol for retrieving and recomposing XML and HTML documents stored on a remote server [12]. DARC provides a contextual search and retrieval capability useful for lightweight web applications such as AJAX or PHP middleware that removes the need for schemas and schema management while providing the capability to retrieve relevant data from large, semi-structured stores.

## 3. METHODOLOGY

A key feature of the XDB system is that this system doesn't require any formally defined semantics (schema) of the data in the documents. Structure and semantics information implicit in the document is exploited instead. This leads to a system where sophisticated data integration and composition applications can be built without high schema management overheads, in a highly scalable manner [13]. Figure 1 shows the high level view of the XDB system which mainly consists of a Web interface with APIs, the XDB Database, and the XML parser.

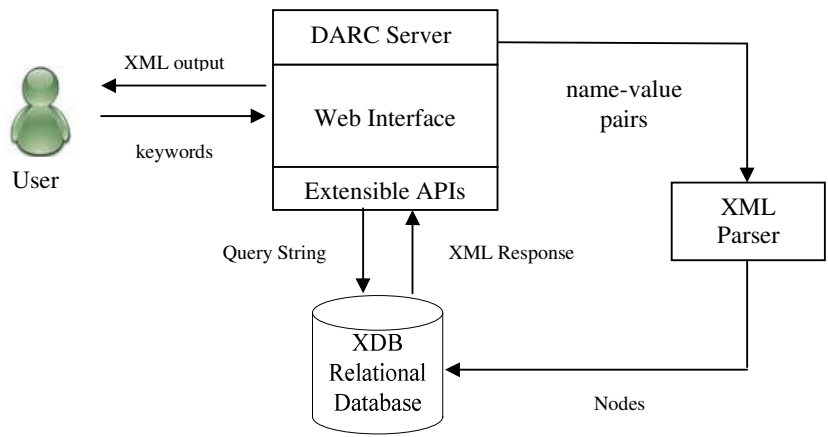

Figure 1. XDB System Overview

The XDB stores information of resources in heterogeneous formats, organizes resources in hierarchy, and allows context and content querying of information in the database. The data querying capabilities in XDB are centered around the notions of context and content in documents treating a document as a collection of various sections and sub-sections opposed to the conventional keyword search approach which treats a document as a single unit. A context is essentially a section or subsection within a document. The information within a context is referred to as content. When a user enters the keywords for context and content to search for information, a query with one or more expressions, separated with the ampersand ("&") character, is generated.    As an example, the query for name "Richard" can be written as http://localhost/DARC?context=Name&content=Richard. After system searches and parses the documents, contexts that match any of the expressions are returned and displayed in XML format on the web browser.

Current XDB protocol, however, defined and implemented for documents in XML and HTML formats only.  To extend the XDB for more and different document formats, we selected Wiki, a new document format, and developed a prototype of extended XDB by using XDB as a reference implementation [14]. The details of XDB and extended XDB implementation are described in following subsections.

## 2.1. Specification of DARC

The Data Access and Retrieval Composition (DARC) protocol provides contextual search and retrieval of semi-structured data. Contextual search is searching for data based on where it is located in the structure, primarily focused on the immediate context of the information sought. For example, if a user wishes to find the information in the "name" section of a set of documents, the search can be quickly performed with no fore-knowledge of the structure of the documents being searched. Contextual retrieval adds the capability to retrieve only the relevant section of the documents that match the query expression, which reduces the network burden required to find the relevant information and provides the opportunity for further analysis. Following the previous example of "name" section, the user's query returns the documents containing the specified information in the "name" section in a structured output. The user could then store the output for further analysis.

DARC is a protocol that follows the Representational State Transfer (REST) architecture. DARC queries are performed as HTTP's GET or POST method, and DARC query results returned in the HTTP response in XML or ASCII. Queries are read-only, stateless operations on the DARC data store. DARC uses standard HTTP requests, caching, proxying, authentication and access control, encryption, compression and other HTTP-related technologies. To update data in a DARC index, WebDAV was used to leverage the simplicity of remote data management. DARC relies on documents having Unique Resource Identifiers (URI) both in the query expressions and in the result output when it identifies matches. DARC is aware of any updates to the underlying resources and each URI uniquely identifies one and only one resource as per the semantics of the URI standard. All DARC queries, as Restful operations, require a base URI. All query expressions and options are specified as part of the "query string" of the URI and the query response is an XML HTTP response body so that users can interact with a DARC system directly from within their browser by typing queries into the URL field and examining the structured XML in the page view of the browser. For an example, a GET request of http://pub.library.org/darc?name=Yoo&syntax=xml would return:

```
<?xml version="1.0"?>
<resultset>
<result>
<meta><uri>/specifications/darc.xml</uri></meta>
<value><name>Yoo</name></value>
</result>
</resultset>
```

Contextual searches support a number of full-text operations on the contents such as stemming, Boolean operators, etc. Contextual searches may be performed using a number of expressions (a DARC server must support all variants of the context search):

- context={content}&content={content}

- {context}={content}

- {context}_co_{content}

- term={context}={content}

- {context}_nc_{content}          // negation operation

In addition, the content can be omitted to search for contexts that match no matter what the content.

The default output of a DARC query is an XML response for each resource that matches and, by default, includes the relevant context and content from the source in a simple XML structure. The output complies with the XML schema. For example, a result for the query of "name=Yoo" is displayed in Figure 2.

```
<?xml version="1.0"?>
<resultset><query>
          <uri>http://localhost/darc?name=Yoo</uri>
          <depth>100</depth>
          <maxhits>10000</maxhits>
          <syntax type="xdb"/>
          <time>Tue Aug 24 11:21:15 2010</time>
          <version>$Revision: 1770 $</version>
</query>
<result><meta>
          <uri>http://localhost/data/output.xml</uri>
          <lastprocessed>2010-08-24T11:20:52-0800</lastprocessed>
</meta><value>
          <context>
                    name
                    <content>Wook-Sung Yoo</content>
          </context>
</value></result>
<result><meta...
</result>
...
</resultset>
```

Figure 2.  An example of output result for the query of "name=Yoo"

DARC servers return identical results to the expressions which are syntactically different but semantically equivalent.

## 2.2. Prototype Development with Wiki

The XDB system uses XSLT to display the results of the query in the browser and the output could either be in XML or ASCII format. We expanded the XDB to Wiki document to facilitate the same features for other document format. Wiki markup is a lightweight markup language used to write pages in Wiki websites, such as Wikipedia, and is a simplified alternative/intermediate to HTML. The content and context search feature in Wiki prototype enables an end user to retrieve information from highly complex and constantly changing heterogeneous data formats into a well-structured, common standard so that the end-user can index documents in Wiki text format by running a single command. The extended XDB system allows the end-user to index documents in Wiki text format by running a single command and DARC queries are executed as HTTP's GET requests and DARC query results are returned in the HTTP response in XML. As described in Figure 3, the prototype handles the query with two main phases: (1) Wiki to XML conversion using SGML, and (2) indexing the converted XML in the database. PerlScript is used to insert converted document to database [15].

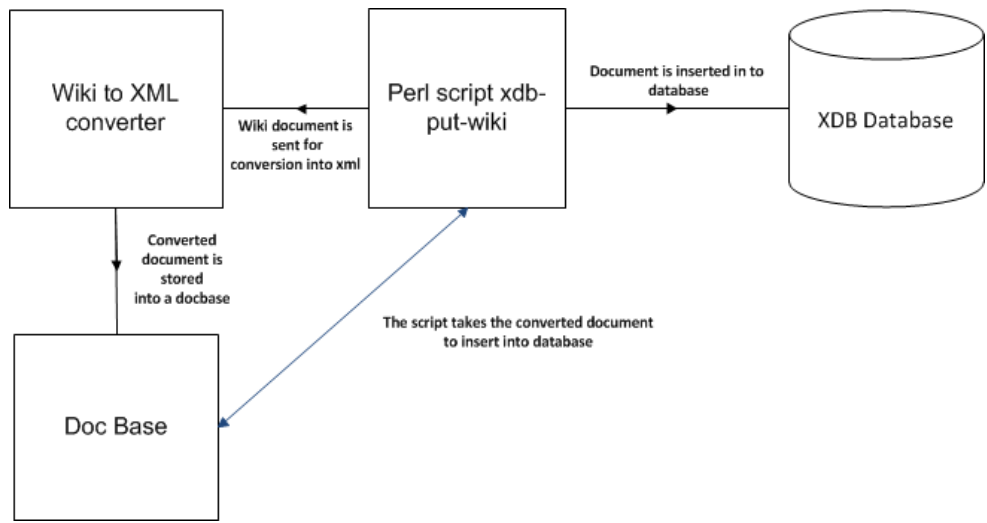

Figure 3: Diagram for the Prototype of Extended XDB

The Wiki to XML Converter Module has two sub-modules: SX (SGML to XML) and Wiki.spm (DTD) modules. The SX module, a free object-oriented toolkit for SGML parsing and entity management, converts SGML to XML. The SX module parses and validates the SGML document and writes an equivalent XML document to the standard output. The SX module sends a warning for SGML constructs which have no XML equivalent. SX takes the following syntax with arguments:

*sx [ -Cehilprvx ] [ -bencoding ] [ -ccatalog_file ] [ -Ddirectory ] [ -ffile ]*
*    [ -wwarning_type ] [ -xxml_output_option ] sysid...*

*    -bencoding*
*            Use encoding for output. By default SX uses UTF-8.*
*    -cfile*
*            Use the catalog entry file.*
*    -C*
*            This has the same effect as in nsgmls.*
*    -Ddirectory*

> *Search directory for files specified in system identifiers.*
> *This has the same effect as in nsgmls.*

-*e*

> *Describe open entities in error messages.*

-*ffile*

> *Redirect errors to file.*
> *This is useful mainly with shells that do not support redirection of stderr.*

-*iname*

> *This has the same effect as in nsgmls.*

-*v*

> *Print the version number.*

-*wtype*

> *Control warnings and errors according to type.*
> *This has the same effect as in nsgmls.*

-*xxml_output_option*
> *Control the XML output according to the value of* xml_output_option

The Wiki.SGML (DTD) Module is used to convert Wiki to SGML. DTD in SGML is used to convert Wiki to SGML by defining a character set including start and ending tags and mapping Wiki tags to corresponding XML. Figure 4 shows 16 tags converted.

| wikitag | What it means | | wikitag | What it means |
|---------|---------------|---|---------|---------------|
| ---- | Horizontal line | | ---...--- | Third level heading |
| * | Bulleted | | ====...==== | Fourth level heading |
| ** | Bulleted(second level) | | [[ ]] | hyperlinks |
| *** | Bulleted(third level) | | '...'' | Italic |
| # | Numbered | | '''...''' | Bold |
| ; | Definition list | | '''''...''''' | Bold+Italic |
| -...- | 1st level heading | | Blank line | New paragraph |
| --...-- | 2nd level heading | | ## | Second level number |

Figure 4.  Converting Wiki to XML in 16 tags

A PerlScript, *xdb-put-wiki*, is used to put the searched document to XDB database by using the following command:

> *xdb-put-wiki  [database name]  [URI]   [filepath]*

The format is consistent with *xdb-put-xml* and *xdb-put-html* in XDB. All source files are saved in */test/data* directory and all error logs is saved in */error* directory.

Figure 5 shows a sample document converted by the PerlScript, *xdb-put-wiki*, from Wiki format to XML format.

Figure 5. Sample Documents Converted from Wiki to XML Format

## 4. CONCLUSIONS

With the majority of today's business contents in semi-structured format, the effective integration of data across an organization without pre-arranged conversion became important to the organization. An extensible database (XDB) enables quick and automatic linking of diverse organizational contents to accelerate the processes and productivity of the organization. Fairfield project team proved the concept of XDB to wiki documents by developing the prototype of extended XDB successfully. The updated XDB protocol and source code of prototype were posted as open source at Source Forge (sourceforge.net). The conversion process of the specification document to MediaWiki is also posted at Source Forge and updated information was shared via WikiPage. The SGML DTD currently supports 16 main tags (MediaWiki). Future development includes development of all tags of MediaWiki and extending to other dialects of Wiki.

## ACKNOWLEDGEMENTS

This research was supported by the National Space Grant Foundation (NASA Technology Database: AERC2-05-SD). Special thanks to Dr. David Maluf and Chris Knight at NASA Ames Research Center and Spoorthy Gowda, Vishwanath Mamillapalli, and other Fairfield University students involved in the project.

## REFERENCES

[1] Halevy, A. & Ashisi, N. & Bitton D. & Carey, M. & Draper, D. & Pollock, J. &. Rosenthal, A., Sikka, V. (2005) "Enterprise information integration: successes, challenges and controversies", *Proceedings of the 2005 ACM SIGMOD international conference on Management of data (SIGMOD '05)*, pp. 778-787.

[2] Moore, C., (2002) "Diving into Data", *InfoWorld*.

[3] Maluf, D. A. & Bell, D. & Ashish, N. & Knight, C & Tran, P. B., (2005) "Semi-Structured Data Management in the Enterprise: A Nimble, High-Throughput, and Scalable Approach", *Proceedings of the 9th International Database Engineering & Application Symposium (IDEAS'05)*, pp.115-124.

[4] Maluf, D. A. & Tran, P. B. & Gawdiak, Y, (2002) "New Information System Provides Multiple Data Source Access", NASA Ames Astrogram Article.

[5] Maluf, D. A. & Tran, P. B., (2001) "Articulation Management for Intelligent Integration of Information", IEEE Transactions on Systems, Man, and Cybernetics, November 2001, vol. 31, No. 4, pp. 485-496.

[6] Maluf, D.A. & Bell, D. G. & Knight, C. & Tram, , P.& La, T. & Lin, J & McDermott, B., (2003) "XDB-IPG: An Extensible Database Architecture for an Information Grid of Heterogeneous and Distributed Information Resources", Stanford Article, Information Management: XDB-IPG-0.9.

[7] Network Working Group of the Internet Engineering Task Force (1999), *HTTP Extensions for Web Distributed Authoring and Versioning (WebDAV)*. RFC 2518.

[8] Harold, E. R. (1998) *XML: Extensible Markup Language*, IDG Books Worldwide.

[9] Bray, T., Paoli, J. & and Sperberg-McQueen, C. M. (1998) "Extensible Markup Language (XML)", World Wide Web Consortium Recommendation REC-xml-19980210, February, 1998.

[10] WebDAV Resources (2010), http://www.webdav.org/.

[11] NASA Information Power Grid (IPG). http://www.ipg.nasa.gov/.

[12] Whitehead, Jr., J. E., & Goland Y. Y., (1999) "WebDAV: a network protocol for remote collaborative authoring on the Web", *Proceedings of the sixth conference on European Conference on Computer Supported Cooperative Work (ECSCW'99)*.

[13] Xerox DocuShare - DocuShare CPX XDB (2007), *The Xerox DocuShare CPX Extensible Database - Real Time Connection of XML Content*. White Paper.

[14] Yoo, W. & Gowda, S. & Mamillapalli V., (2012) "Development of Extended XDB Protocol and Prototype", *Proceedings of the International Conference on Education and Management Technology (ICEMT'12),* pp. 142-145.

[15] Brown, M. C., (2001) *XML Processing with Perl, Python, and PHP*, Sybex Inc.

**Author:**

Wook-Sung Yoo, Ph.D. is a chair and an associate professor in the Software Engineering department at Fairfield University. Prior to joining Fairfield University, he was an associate professor at Computer and Information Science Program, a director of Bioinformatics program, at Gannon University, PA. He also worked at the Oral Health Service and Informatics Department at the SUNY at Buffalo. He was a chair of the IMIA (International Medical Informatics Association) Dental Informatics working group and involved in various national wide web and informatics projects. 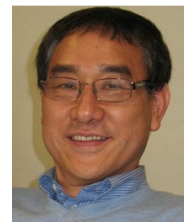